\documentclass[12pt,preprint]{aastex}

\usepackage{epsfig}

\newcommand{\avg}[1]{\langle{#1}\rangle}

\newcommand{\ltsima}{$\; \buildrel < \over \sim \;$}
\newcommand{\lsim}{\lower.5ex\hbox{\ltsima}}
\newcommand{\gtsima}{$\; \buildrel > \over \sim \;$}
\newcommand{\gsim}{\lower.5ex\hbox{\gtsima}}

\def\gtrsim{\mathrel{\hbox{\rlap{\hbox{\lower4pt\hbox{$\sim$}}}\hbox{$>$}}}}

\def\lesssim{\mathrel{\hbox{\rlap{\hbox{\lower4pt\hbox{$\sim$}}}\hbox{$<$}}}}

\begin{document}
\title{Measuring the Three Point Correlation Function of the Cosmic Microwave
Background}
 
\author{Gang Chen\altaffilmark{1} and Istv\'an Szapudi\altaffilmark{1}}
 
\altaffiltext{1}{Institute for Astronomy, University of Hawaii,
2680 Woodlawn Dr, Honolulu, HI 96822, USA}

\begin{abstract}

We present a new method to estimate three-point correlations
in Cosmic Microwave Background maps. Our Fast Fourier Transform 
based implementation 
estimates three-point functions using
all possible configurations (triangles) at a controlled
resolution. The speed of the technique depends both on the
resolution and the total number of pixels $N$.
The resulting $N\log N$ scaling is substantially faster
than naive methods with prohibitive $N^3$ scaling. As an initial application, 
we measure three-point correlation functions in the First Year
Data Release of the Wilkinson Anisotropy Probe.
We estimate 336 cross-correlations of any
triplet of maps from the 8 differential assemblies,
scanning altogether 2.6 million triangular configurations.
Estimating covariances from  Gaussian signal plus realistic noise
simulations, we perform a null-hypothesis testing
with regards to the Gaussianity of the Cosmic Microwave Background.
Our main result is that at the three-point
level WMAP is fully consistent with
Gaussianity. To quantify the level of possible deviations,
we introduce false discovery rate analysis, 
a novel statistical technique to analyze for three-point measurements.
This confirms that the
data are consistent with Gaussianity at better than 1-$\sigma$ level 
when  jointly considering all configurations.
We constrain a specific non-Gaussian model using
the quadratic approximation of weak non-Gaussianities
in terms of the $f_{NLT}$ parameter, for which we construct
an estimator from the the three-point function. We
find that using the skewness alone is more constraining
than a heuristic suboptimal combination of all our results;
our best estimate is $f_{NLT} = -110\pm 150$ assuming a
$\Lambda$CDM concordance model.

\end{abstract}




\section{Introduction}

The temperature fluctuations in the Cosmic Microwave Background  (CMB) 
are Gaussian to a high degree of accuracy
\citep{KomatsuEtal2003a}. Non-Gaussianity, if any, enters
at a highly subdominant level. It could be either primordially
generated along with Gaussian fluctuations by 
exotic inflationary models, and/or it could arise from secondary anisotropies,
such as gravitational lensing, Sunyaev-Zel'dovich (SZ),
or Sachs-Wolfe (SW) effects. Quantifying the
degree and nature of non-Gaussianity in the CMB
constrains specific inflationary models, as well
as enhances our understanding of the secondary processes
the CMB underwent beyond the surface of last scattering.
Interpretation of any such measurement is complicated by the
fact that systematics and foreground contaminations might also
produce non-Gaussian signatures. 

Given the nearly Gaussian nature of the CMB, 
$N$-point correlation functions, and their harmonic counterparts,
polyspectra,  are the most natural tools for
the perturbative understanding of non-Gaussianity.
If it were generated by inflationary models admitting a $\Phi^2$ term, 
the leading order effect would be a $3$-point function. On the other
hand some secondary anisotropies, such as lensing, 
are known to produce 4-point non-Gaussianity at 
leading order \citep{Bernardeau1997}.
The skewness (or integrated bispectrum) was measured by 
\cite{KomatsuEtal2003a} and $3$-point correlation 
function by \cite{GaztanagaWagg2003, EriksenEtal2005}.

Many alternative statistics have been used to investigate 
non-Gaussianity in CMB. A partial list includes wavelet
coefficients \citep{VielvaEtal2004, MukherjeeWang2004, LiuZhang2005, McEwenEtal2005}, 
Minkowski functionals \citep{KomatsuEtal2003a, Park2004}, 
phase correlations between spherical harmonic 
coefficients \citep{NaselskyEtal2005},
multipole alignment 
statistics \citep{deOliveira-CostaEtal2004, CopiEtal2004, SlosarSeljak2004}, 
statistics of hot and cold spots \citep{LarsonWandelt2005, TojeiroEtal2005, 
CruzEtal2005},
higher criticism statistic of pixel values directly \citep{CayonEtal2005}.
Most of these measurements are consistent with Gaussianity,
although some claim detections of  non-Gaussianity 
up to 3-$\sigma$ level. These 
alternative statistics, albeit often easier to measure, typically
depend on $N$-point functions in a complex way, thus
they cannot pin-point as precisely the source of non-Gaussianity.

Among the three-point statistics, there is a perceived complementarity
between harmonic and real space methods.
The bispectrum can be relatively easily calculated for a full
sky map \citep{KomatsuEtal2002}, although the present methods have
a somewhat slow $N^{5/2}$ scaling \citep{KomatsuEtal2003b}. Methods
put forward so far use the ``pseudo-bispectrum'', ignoring
the convolution with the complicated geometry induced by 
galactic cut and cut-out holes. In contrast with harmonic
space, the corresponding
pixel space edge  effect corrections are trivial \citep{SzapudiEtal2001},
since the window function is diagonal. Unfortunately, 
simple methods to measure three-point clustering exhibit
a prohibitive $N^3$ scaling if the full configuration space is
scanned. To remedy the situation, most previous
measurements of the $3$-point
function only deal with an ad-hoc sub-set of triangular configurations
\citep{GaztanagaWagg2003, EriksenEtal2005}. Both of these papers
covered the full configuration space on small scales; the former
paper also appears to have estimated most  configurations 
on large scales, missing intermediate configurations
with mixes scales.

This work presents a novel method, 
which, at a given resolution, scans
the full available configuration space for $3$-point level statistics
using realistic computational resources.
We find that the resulting configuration
space itself is overwhelming to such a degree that interpretation
of the results also requires novel methods. We 
introduce false discovery rate (FDR) technique 
as a tool to interpret three-point correlation function measurements.

The next section introduces our algorithm to measure the
$3$-point correlation function, \S 3
illustrates it with an application to 
the WMAP first year data release,
and \S 4 introduces the FDR method and applies it
to our results. We summarize and discuss our results in \S 5.

\section{Measuring the three point correlation function}

The three point correlation function \citep[e.g.,][]{Peebles1980}
is defined as a joint moment of three density fields
$\zeta = \avg{\delta_0\delta_1\delta_2}$ at three spatial positions.
For CMB studies $\delta_i$ denotes temperature fluctuations at position $i$
on the sky, and $\avg{}$ stands for ensemble average. If the underlying
distribution is spatially isotropic, $\zeta$ will only depend on the
shape and size of a (spherical) triangle arising from the three
positions. A number of characterizations of this triangle are possible
and convenient. The most widely used are
the sizes of its sides (measured in radians), or two sizes and
the angle between them. This latter angle is measured on
the spherical surface of the sky.

One can use the ergodic principle of replacing ensemble averages
with spatial averages to construct a nearly optimal, edge corrected estimators
with heuristic weights 
\citep{SzapudiSzalay1998,SzapudiEtal2001,SzapudiEtal2005}
\begin{equation}
\zeta(\Delta) = \frac{
\sum_{i,j,k}f^\Delta_{i,j,k} \delta_i\delta_j\delta_k w_i w_j w_k}{
\sum_{i,j,k}f^\Delta_{i,j,k} w_i w_j w_k},
\label{eq:estimator}
\end{equation}
where we symbolically denoted a particular triangular configuration
with $\Delta$ (any parametrization would suffice), 
and $f^\Delta_{i,j,k} \propto 1$
if pixels $(i,j,k)\in \Delta$, and $0$ otherwise. We also defined
a multiplicative weight $w_i$ for each pixel: this is $0$ if
a pixel is masked out, and it could take various convenient values
depending on our noise weighting scheme if the pixel is inside the
survey; e.g., in the case of flat weights it is simply $1$.
This simple estimator has been widely used in large scale structure,
and it is nearly optimal with appropriate weights.
\citep[e.g.,][]{SzapudiSzalay1998,KayoEtal2004}. It
is entirely  analogous to the successful estimators used 
for the measurements of the $C_l$'s for the CMB
\citep[up to harmonic transform,][]{SzapudiEtal2001,HivonEtal2002}.

The naive realization of Equation~\ref{eq:estimator} has
a prohibitive  $N^3$ scaling
if one needs to scan through triplets of pixels and assign them to
a particular bin. The summation can be restricted and thus
made faster if one restricts the number of configurations and the
resolution
\citep[e.g.,][]{SzapudiEtal1999b,BarrigaGaztanaga2002,GaztanagaWagg2003},
or it can be sped up by using tree-data structures \citep{MooreEtal2001}.
Neither of these methods is able to scan through all possible configurations
in megapixel maps
with reasonable amount of computing resources. Here we propose a new method
which uses both hierarchical pixelization and Fourier methods 
motivated by \cite{Szapudi2004,SzapudiEtal2005} to scan through all
the triangles simultaneously. Note that \cite{GaztanagaWagg2003} comes
closest to our aims, but their simple two-step approach is not systematic
enough to cover all possible triangles at a given resolution, and
it is not fast enough for massive Monte Carlo simulations.

In the following we will choose a parametrization of the triangle $\Delta$
using two of its sides $\theta_1,\theta_2$, and the angle $\alpha$ 
between them. We define the configuration
space as a set of (logarithmic) bins for the sides, and linear bins for
the angle in their full possible range, i.e., $0,\pi$ (remember that
the sides of the triangle on the sky are also measured in radians).
The given resolution is determined by the number of bins for $\theta_i$, and
the number of bins for $\alpha$.
Note that a particular triangle might appear more than once in this
scheme, albeit with different resolutions.
 Different triangular bins of the three-point function
are strongly correlated anyway, and the correlation from
duplicating triangles can be taken
into account in the general statistical framework 
over correlated bins.

Given a triangular configuration, 
and a pixel $i$, all other pixels which enter the
summation in Equation~\ref{eq:estimator} are located on two
concentric rings of size $\theta_1$ and $\theta_2$. As a consequence,
the summation over fixed $\alpha$ can be thought of as an unnormalized
(raw) two-point correlation 
function between  two rings. To obtain three-point correlation
function, one has to multiply this two-point correlation function
with the value of the center pixel $i$ and finally sum over $i$.

\begin{figure}[htb]
\epsscale{1.}
\plotone{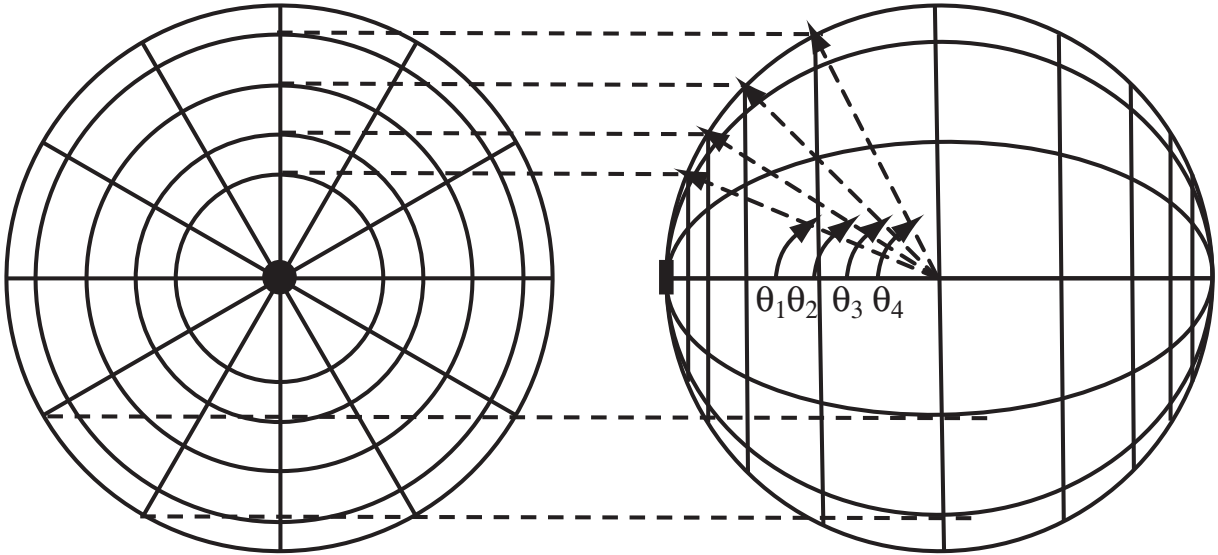}
\caption{The repixellization geometry viewed from position 
above the center and from 
the side of the center.  
$\theta_i$s on the side view define circles on the top view 
that separate the rings. The radius on 
the top view corresponding to big circles on the side view that cut rings to 
the new ring-pixels.}
\label{fig:rings}
\end{figure}

Calculating the two-point correlation function of rings can be fast
if one repixellizes the map (c.f., Fig.~\ref{fig:rings}) into
rings with sizes matching the binning scheme for $\theta$,
and uniform division in $\alpha$. Such a repixellization, 
resulting in ring-pixels as shown in Fig.~\ref{fig:rings},
would take only $N$ steps even in
a naive way; the HEALPix
hierarchical scheme allows it to be done in $\log N$ time.

We use the following algorithm: let us 
start a recursive tree walk at the coarsest map,
$N_{side}=1$ in the HEALPix scheme. For each pixel in this map,
we determine, using its center, which ring-pixel it would belong to.
If the size of the pixel is much smaller than this ring-pixel (how much
smaller is a parameter of our algorithm: in this paper we used
the condition that the pixel has to be smaller then $0.2\times$
the bin width which is also the approximate size of the ring-pixels), 
we record it. If not, algorithm splits the quad-tree,
and calls itself recursively for each four sub-pixels. This procedure
ends at the latest when the highest resolution (i.e. the one of the
underlying map) is reached. If the
bins are chosen appropriately such that large ring-pixels are set up for large
triangles, for many pixels it will finish earlier. As noted above,
the map has to be regridded around each pixel into rings of ring-pixels. In total,
this takes $O(N\log N)$ time. 

Calculating the two-point correlation function between rings 
speeds up using Fast Fourier Transform (FFT) methods, such as
those put forward in \cite{SzapudiEtal2001,SzapudiEtal2005,Szapudi2005}.
The recipe is the following. First, FFT every ring $i$ to obtain 
complex coefficients $a_k(\theta_i)$; then calculate for every
pair of rings $(i,j)$ the ``pseudo power spectrum''
$a_k(\theta_i)a^*_k(\theta_j)+a^*_k(\theta_i)a_k(\theta_j)$, where $^*$ means complex conjugate.
Due to the U(1) symmetry of the ring an inverse $\cos$ transform
will give the (raw) two-point correlation function between the two
rings \citep[c.f.][]{SzapudiEtal2005,Szapudi2005}. 
If we have $N_\theta$ rings, each of them 
$N_\alpha$ ring-pixels, each FFT can be done in 
$N_\alpha \log N_\alpha$ time, and there is $N_\theta(N_\theta+1)/2$
cross correlations to be calculated for a full scan of configurations.
All the above needs to be performed for each pixel 
as a center point.
The total scaling (including the initial regridding) takes
$N(\log N+N_\theta N_\alpha \log N_\alpha + 
N_\alpha N_\theta(N_\theta+1)/4)/2$,
where we took into account that the two opposite pixels can be 
handled in one go if a symmetric set of bins around $\pi/2$ 
are used for $\theta$. 

While the above procedure to calculate raw (unnormalized) 
correlation functions appears somewhat complex, we have checked
with direct calculation that it gives numerically the same
result as calculating correlations on the rings in a naive way.
In order to obtain normalized correlation functions, the
same procedure has to be followed for the rings associated with
weights/masks. Each configuration of the raw three-point function
is divided with the mask/weight three-point function
for the final result.
For many realizations with the same mask, such as in the case
of massive Monte Carlo simulations, the mask correlations
need to be estimated  only once, representing negligible cost.

The above abstract scheme and calculation
will be illustrated and further clarified 
with a practical application to WMAP next.

\section{Application to WMAP: raw results}

\subsection{Data and Simulations}

We demonstrate 
our method to calculate the three-point correlation function
with an application to WMAP.
We downloaded first year foreground cleaned maps
from LAMBDA website
\footnote{http://lambda.gsfc.nasa.gov/}\citep{BennettEtal2003a}. 
There are total 8 maps for 8 Differencing Assemblies (DA) 
in Q, V and W bands: Q1, 
Q2, V1, V2, W1, W2, W3, and W4, already in HEALPix format. Following
the two-point analysis of WMAP \citep{SpergelEtal2003, FosalbaSzapudi2004},
we only used cross correlations, i.e. three-point correlation
functions calculated from three different DAs. 

We produced 100 (Gaussian) simulations with
SYNFAST in HEALPix package\footnote{http://www.eso.org/science/healpix/}. 
The input power spectrum, 
also from the LAMBDA website, was taken from $\Lambda$CDM 
model using a scale-dependent (running) 
primordial spectral index which best fits the WMAP, 
CBI and ACBAR CMB data, 
plus the 2dF and Lyman-alpha data. 
Every simulation consists of 8 assembly maps as the data 
\citep{SpergelEtal2003}. 
These 8 maps were generated with a same random seed, 
representing the same primordial CMB,  but 8 different beam transfer functions.
Then different simulated noise maps from LAMBDA web \citep{HinshawEtal2003}
were added to the SYNFAST output maps.

Since the non-Gaussian signal is exceedingly small, and on the smallest 
scales the data are noise dominated, we degraded all maps (simulations and
data) to $N_{side}=256$ after applying the kp2 mask. More precisely,
we added up pixel and weight values for each map; our two
weighting schemes are presented in the next subsection.


\subsection{Binning and Weighting}

At the heart of our algorithm is the regridding of Figure~\ref{fig:rings}
which matches our binning of the triangles.
We chose 19 rings for half of the sphere surface, and the same
bins are repeated on the other half symmetrically around $\pi/2$.
The 19 bins are chosen to be uniformly distributed in logarithm
between $\frac{\pi}{2}/n_{side}$ and $\frac{\pi}{2}$.  
The number 19 was chosen such that $\theta_{i+1}^2 \simeq 2\theta_i^2 $,
which gives a logarithmic resolution of $\sqrt{2}$.
Every ring was divided to 20 ring-pixels in $\alpha$. This number
renders the resulting ring-pixels fairly compact, 
and it is also convenient for our chosen implementation of FFT
\citep[FFTW,][]{FrigoEtal2005}. 

Weight maps were constructed using the kp2 mask and the noise
profile of the maps.
We used two weighting schemes: flat 
weighting where $w_i$ is 1 or 0 depending on the mask,
and (inverse) noise weighting; for the latter \citep{BennettEtal2003a} we used
the effective number of observations of the pixel.
The weights need to be determined only up to  multiplicative factor, 
as their overall normalization cancels from the algorithm.
The average noise level $\sigma_0$ for each DA 
is used when combining $\zeta$ over different cross correlations.

The total number of triangular configurations in $38$ rings
with $11$ possible values of $\alpha$ (angles large than $\pi$
count to $2\pi-\alpha$) is $38\times 39/2\times 11=8151$ for
autocorrelations. 
The same number is valid for a cross correlations 
(which we will be exclusively doing) of 3 DAs. We introduce
the notation (DA1, DA2,
DA3), for the central pixel, the first ring and the second ring 
sampled from the three DA in this order. In addition 
we restrict the ``first ring''
has $\theta_i$ no larger than that of the ``second ring''. 
Then the total number of cross correlations between the 8 DA's is 
$8\times7\times6 =336$.
Effectively, each triangle is calculated six
times for a given triplet of  DA's due to the possible 6 permutations.
However, each sample has a different resolution, therefore we
opt to keep all possibilities.
The resulting correlations are taken into account when dealing
with correlated bins in general. In total, there are
about $2.6\times 10^6$ triangular configurations for each data or
simulation set. Note that 
the total number of triplets of ring-pixels examined is $20\times N_{pix}$ 
more, or $4.4\times 10^{13}$. This is still a lot smaller than checking
$10^{18}$ triplets of pixels naively in an $N_{size}=256$ map.
These numbers suggest that our algorithm even without FFT should
take order of days, 
while the naive algorithm would need over 200 years of CPU.

For a batch of $10$ simulations (comprising of $10\times 8$ DA maps),
the calculation of the full three-point function in
all the configuration takes about 90 hours on an Intel Xeon 2.4GHz CPU.
This means each cross-correlation takes only about $2$ minutes on
average! About 10 hours are saved by batch processing 10 sets of simulations:
to estimate $\zeta$ in the data alone (one set of 8 DA files) took 
10 hours.

Clearly, the given resolution does not extract all the information
from the data, as there are approximately $O(N^{3/2})$ distinct 
configurations of the bispectrum or three-point function. However,
it surely must be redundant to extract more configurations than
the amount of data. The ratio of data points vs. configurations is
about $50$ for our chosen bins. It is unlikely that it were
fruitful to push this number towards much smaller values, although
the speed of our algorithm would allow higher resolution.

\subsection{The Three-Point Function of WMAP}

\begin{figure}[htb]
\epsscale{1.}
\plotone{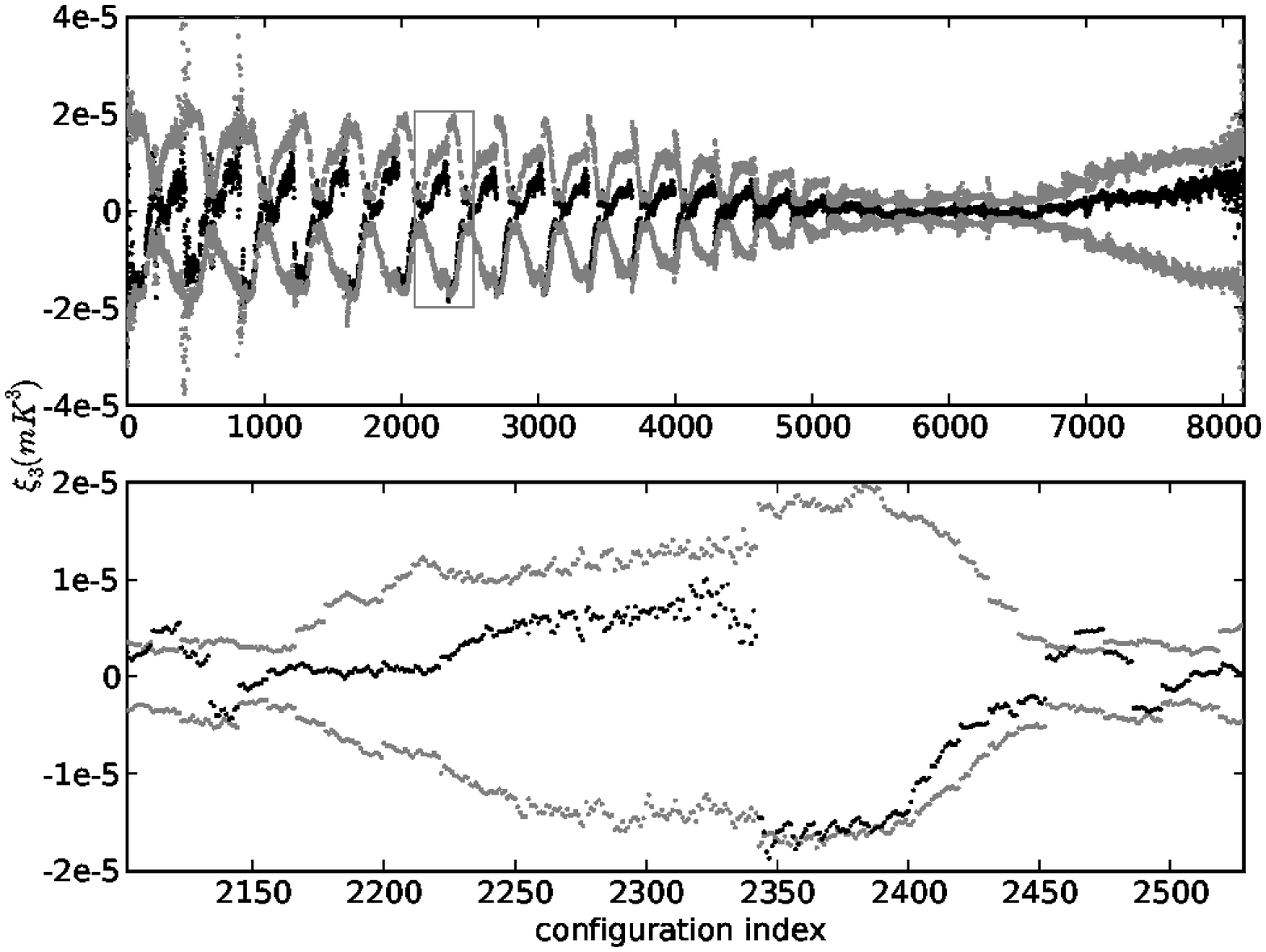}
\caption{$\zeta$ values of all  8151 configurations for
the cross correlation (W2,W3,W4).
Top: the light gray points show the middle 68\% range 
estimated from  100 Gaussian simulations,
the dark gray points correspond to $\zeta$ from WMAP. 
Bottom: zoom on the details inside the small box in the 
top. }
\label{fig:zeta}
\end{figure}

Figure~\ref{fig:zeta} shows a typical set of $\zeta$ 
measurements for the (W2,W3,W4) DA cross-correlation.
The results from WMAP lie comfortably 
in the 68\% range of results from Gaussian 
simulations. Similar results are found for the other 
DA combinations, 
or when all the 24 possible W-only
DA results are averaged. Although different combinations
have different effective beam, full averaging is meaningful
on large scales. We  checked that averaging
all 336 possible combinations is also consistent with
Gaussian. Finally, repeating all the measurement with
noise weighting produced no obvious departure from Gaussianity
either.

\begin{figure}[htb]
\epsscale{1.}
\plotone{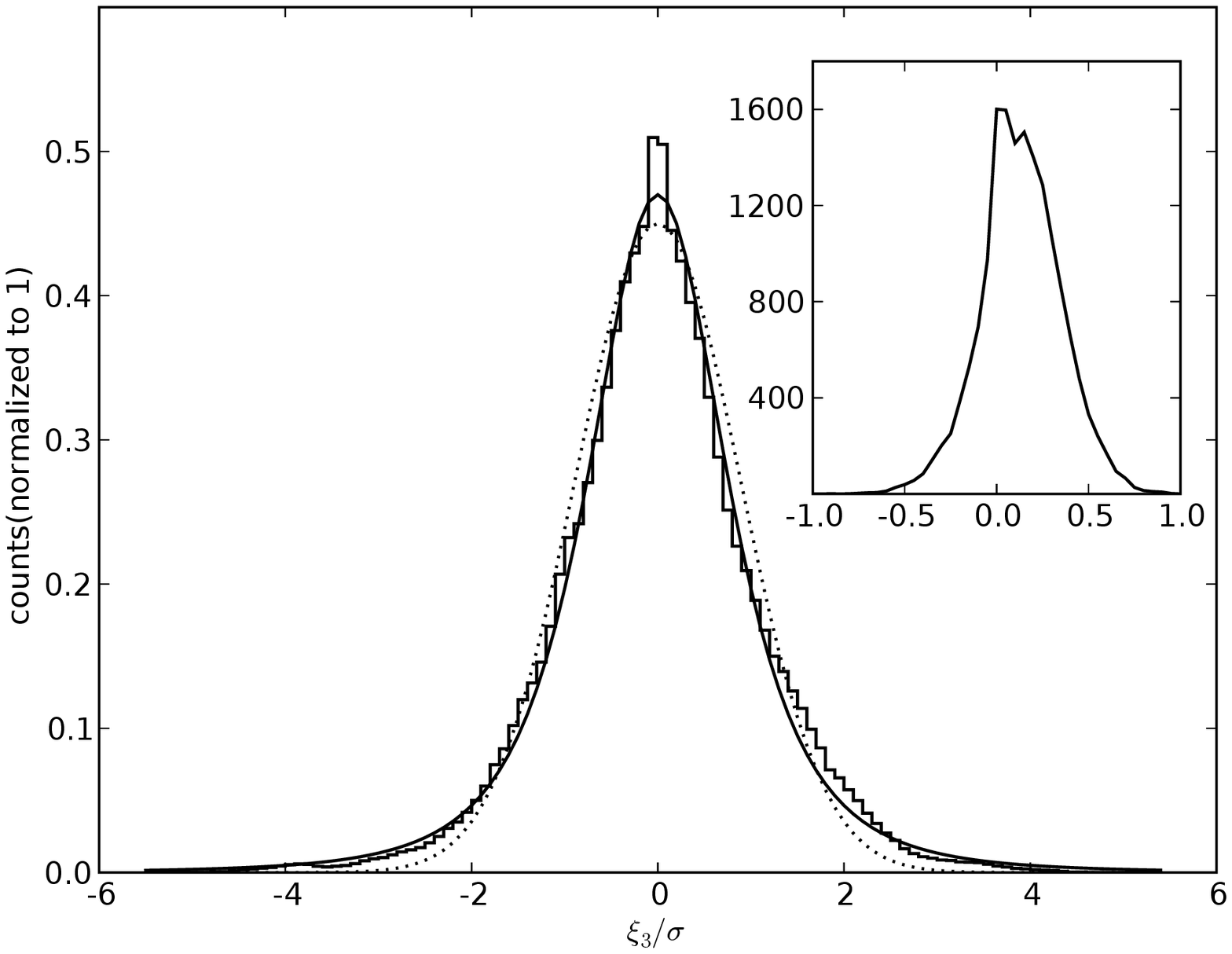}
\caption{Histogram of $\zeta$ values of all 100 simulations for all 
configurations  in (W2, W3, W4). Each value has been
renormalized by subtracting the median, and divided with $\sigma$
calculated from 68 percentiles.
The two curves correspond to best fitting Gaussian (dots) and 
Student function (solid line) with 3 degrees of freedom.
{\em Inset}: histogram of $\chi^2$ differences between fitting
a Gaussian or Student distribution for each bin. According
to the figure, most configurations are better fitted by
a Student distribution as evidenced by the larger $\chi^2$
for the former; however, Gaussian is still a reasonable fit.}
\label{fig:hist}
\end{figure}

At the same time, the scatter in the simulations, i.e., the
probability density function (PDF) of $\zeta$ from 100 simulation,
shows a slightly non-Gaussian signature. For (W2,W3,W4),
Figure~\ref{fig:hist} shows a histogram derived from all 
simulated $\zeta$ values normalized by their measured
median and 68\% levels. Slight deviations from
Gaussian distribution are evident: Student distribution
of degree 3 fits better the overall distribution.
This same distribution produces lower $\chi^2$ when
applied to {\em individual}
triangular configurations according to the inset,
i.e. it is a (marginally) better fit than Gaussian.
We fully take this into account in our hypothesis testing 
which is described next.

\section{Hypothesis Testing}

Our goal is to test the null-hypothesis of Gaussianity against
our measurements by means of comparing the $\zeta$ values measured
from  the data with the corresponding probability distribution
function (PDF) determined from Gaussian simulations.
A crucial step in the traditional 
$\chi^2$ method appears to be computationally infeasible 
due to the large number
of configurations: calculation of the
(pseudo) inverse of an $n\times n$ matrix for $n=2.6\times 10^6$,
the total number of our highly correlated configurations.
Moreover, as seen above, the underlying PDF marginally 
violates Gaussian assumption, even for Gaussian simulations.
Even if it were possible to calculate the inverse of the covariance matrix, 
and we
were to accept the accuracy of the Gaussianity in PDF of the individual
bins, it is not possible to determine the underlying covariance
matrix with sufficient accuracy. In fact, one would need
\citep[e.g.,][]{PanSzapudi2005}
at least (and likely much more than) 2.6 million simulations
for that purpose. 
\cite{LarsonWandelt2005} have shown that
using simulations with uncorrelated noise might result in 
spurious detection of non-Gaussianity.
Therefore we chose to use only the WMAP supplied correlated
noise simulations, of which 110 is available at present.

It is straightforward to test the null-hypothesis with a
single configuration: we can calculate a $p$-value from
the best fit Student distribution from our simulations.
The $p$-value is defined as 
the probability of obtaining a $\zeta$ value that 
is at least as extreme as the one measured from WMAP.
For a threshold $p_t$, the null-hypothesis is rejected 
at $1-p_t$ level if the $p$-value of the datum
is smaller than $p_t$.
A problem arises from combining
2.6 million tests when all the data are
used. For instance,  even if the hypothesis were true, about 260
bins would still be rejected at the $99.99\%$ level (ignoring
the correlations in the data).
 
Fortunately a robust and simple method exists for massive hypothesis
testing, which is insensitive to correlations between the tests,
and makes no assumption on the Gaussianity of the underlying 
error distribution: the method of False Discovery Rate (FDR) 
\citep{BenjaminiHochberg1995, BenjaminiYekutieli2001}.
In astronomy, it has been successfully applied 
in the context of image processing
and finding outliers by \cite{MillerEtal2001}, which 
can be consulted for a more detailed introduction.
The FDR method combines the same $p$-value as defined 
above for individual 
tests using a threshold for rejecting the 
null hypothesis. This combination is insensitive to 
correlations and has more statistical power than naive 
combination.
Our goal is to adapt this powerful method for hypothesis testing of
three-point correlation function measurements with overwhelming
number of configurations.

The FDR method gives a simple prescription for finding a threshold for
rejection. In particular, the recipe suggests that we choose
a threshold such that we control the rate of {\em false rejections}
or FDR. The parameter, taking a similar role to the
confidence interval in more traditional tests, is 
the maximum rate of FDR. If we fix 
an $\alpha$ such that $0\le\alpha\le 1 $,
the FDR procedure will guarantee
\begin{equation}
\avg{FDR} \le \alpha
\label{eq:FDR}
\end{equation}
in ensemble average.

Next we describe the recipe to control FDR; more details can be
found in \cite{MillerEtal2001}. Let $P_1,...,P_{N}$ denote 
the $p$-values calculated from the measurements 
of $N$ configurations, {\em sorted} from smallest to largest. Let
\begin{equation}
d = max\left\{ j:P_j < \frac{j\alpha}{c_N N} \right\}
\end{equation}
where $c_N$ is a constant depending on the level of correlations 
between different configurations.
For uncorrelated data $c_N=1$; while 
$c_N \lesssim \sum_{i=1}^N i^{-1}$ can be used for
correlated data \citep{HopkinsEtal2002}. Note that technically
one would have to adjust $c_N$ to the degree of correlations in
the data.  The suggested value for correlated data is
extremely conservative, and should be considered
as a strong upper limit. Even using this conservative adjustment decreases
the statistical power of the technique only logarithmically;
the final results are expected to be robust
regardless of the degree of correlations.

\begin{figure}[htb]
\epsscale{1.}
\plotone{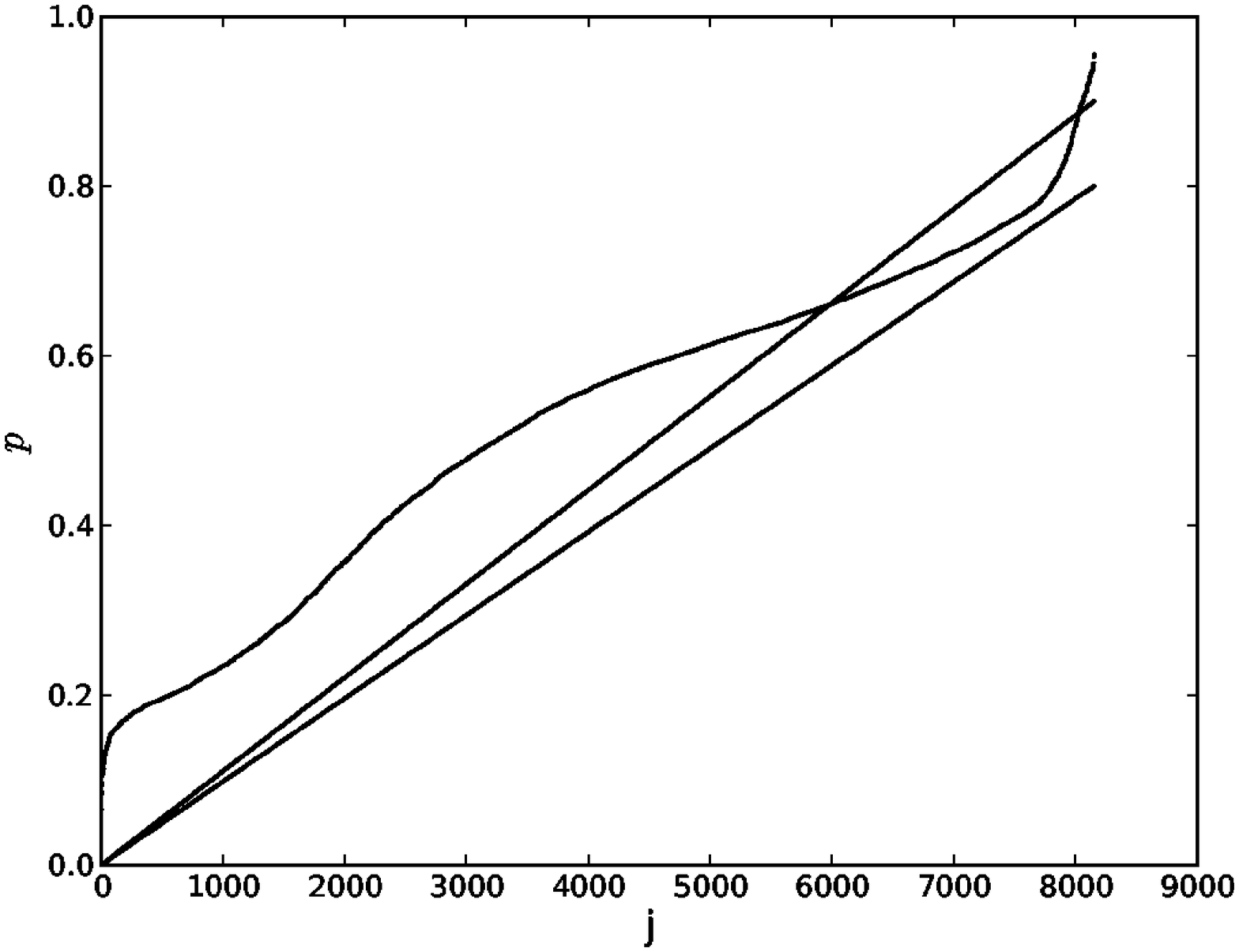}
\caption{The sorted $p$ values vs. lines with slope $\alpha/c_N$.
The $p$ values shown in the curve are for all configurations 
of our typical DA combination $(W2,W3,W4)$.
The lower line is for $\alpha=0.8$ and produces no rejection.
On the other hand, 
the upper line is for $\alpha=0.9$ produces many (8000) rejections,
probably all of them false discoveries (see text).
}
\label{fig:FDR}
\end{figure}

If configurations with $i < d$ are rejected, Equation~\ref{eq:FDR}
will hold \citep{BenjaminiHochberg1995}, i.e. the FDR is controlled according
to our preset parameter $\alpha$. The procedure is
represented graphically on Figure~\ref{fig:FDR}:
$P_j$ is plotted against $j/N$ superposed
with the line through the origin of slope $\alpha/c_N$.
All $p$-values reject the null hypotheses which are to the left
from the last point at which $P_j$ falls below the line.
These might include some false discoveries which are guaranteed to
be a smaller fraction than
$\alpha$ in ensemble average.

\subsection{Application the WMAP $\zeta$}

We have applied the FDR recipe to all of our individual cross-three
point functions, as well as our full data set.
Since $c_N$ is a constant, initially we kept $c_N=1$. 
For a fixed $\alpha$, the results can
be subsequently reinterpreted in terms of any $c_N > 1$. 
For DA combination (W2, W3, W4), 
there is no rejection for $\alpha < 0.81$, i.e., 
allowing as high as 81\% false rejections, 
not a single configuration rejected our Gaussian null hypothesis.
Correlations might increase $c_N$, but it must be $\lesssim 14$. 
The true $\alpha$,
when correlations are taking into account, can only be larger
than our effective $\alpha$ for $c_N=1$. In other words, 
the data are fully consistent with Gaussianity.

As a sanity check, we repeated the FDR analysis in our simulations
as well. By scanning through different $\alpha$ values from 
0 to 1, we find that 50 out of 100 simulations  have rejections with 
$\alpha < 0.81$. This means that the WMAP measurements are 
fully consistent with Gaussianity at a level better than 1-$\sigma$ 
in the traditional sense.
In summary, at the three-point level, scanning all configurations,
we did not find any significant non-Gaussianty which would be
localized in pixel space triangular configurations.

We performed FDR analysis on all $336$ measurements individually, 
as well as on the combination of all those measurements with $2.6$ million
configurations in total. None of these cases produced credible evidence
for non-Gaussianity and all of them were fully consistent with
our null hypothesis at $\alpha \sim 0.8$.

\section{Summary and Discussions}

\subsection{Summary}

We presented a new method to measure angular three-point correlation
functions on spherical maps. We achieve an unprecedented $N\log N$ scaling 
with a combination of hierarchical and Fourier algorithms.
The speed of our technique allows a systematical scan of 
the full available configuration space
at a given resolution. Such speed is especially useful for cross
correlations and Monte Carlo simulations, where a vast number
of configurations and measurements need to be performed.
We have achieved a speed of about $2$ minutes per 
cross correlations, when 336 cross-correlations have been
estimated simultaneously in $N_{side}=256$ HEALPix maps
using a single Intel Xeon 2.4GHz CPU. This is to be contrasted with
a naive approach, which would have taken about 200 years
per cross-correlations; a $20$ million fold speed up.

As a first application of our code we analyzed the first
year WMAP data along with 100 realistic simulations.
We have calculated cross-correlations
for about  $2.6\times 10^{6}$ triangular configurations,
or about $4.4\times 10^{13}$ triplets in total in
maps of $N_{side}=256$ corresponding the $8$ DA's. 
The ratio of pixels/configurations is about 50 for each measurement. 

Comparing our measurements from 100 Gaussian simulations with realistic 
correlated noise, we found WMAP to be comfortably within
the 68\% percent range for most configurations. Any significant
departure from Gaussianity at the three-point level, even if
localized in particular triangular configurations, 
would have shown up clearly in our full scan of the 
available configuration space. Our main
result is that there is no credible evidence of non-Gaussianity at the
three-point level at any of the triangular configurations
we examined. As a consequence, if the tentative detection of
non-Gaussianity claimed in previous works holds up, 
it should correspond to either 4-point or higher order correlations, or
to spatially localized features which break rotational invariance
\citep[e.g.,][]{McEwenEtal2005, CayonEtal2005}. In contrast
with our measurements, 
all previous studies of higher order statistics 
used autocorrelations. Comparison of our errorbars with
that of \cite{GaztanagaWagg2003} appears to show that this
increases the errors by a factor of two (see the discussions below).
In addition, many measurements used uncorrelated
noise simulations. 
According to the findings of \cite{LarsonWandelt2005},
this might increase the likelihood of 
finding spurious non-Gaussianity.

Analysis of our Gaussian simulations revealed that there is a slight
non-Gaussianity in the error distribution of individual configurations.
This is not surprising, since
three-point correlation function is a non-linear construction
of the Gaussian random variables \citep[c.f.][]{SzapudiEtal2000}.
The error distribution is well
fit by a Student distribution with 3 degrees of freedom. 

To quantify any possible departure of the overall data set
from Gaussianity, we
introduced a new technique, FDR, to interpret three-point
statistics. This corresponds to an optimized multiple hypothesis
testing, and it is insensitive to the unavoidable correlations in the
data. All of our FDR tests, whether applied to any of the 336
cross correlations, or the combined data set, were fully consistent
with Gaussianity with better then 1-$\sigma$.
This quantifies our previous assertion based on examination
of the individual configurations under the assumption of statistical
isotropy.

\subsection{Discussions: constraining specific models}

The above model independent tests showed that there is no
credible evidence of any non-Gaussianity in the data. Next we
illustrate how our measurements  yield constraints on specific
non-Gaussian models. We choose
a simple phenomenological model corresponding to the
quadratic expansion of the density field in terms of 
one parameter, $f_{NLT}$, as put forward by
\citep{GaztanagaWagg2003}: 
\begin{equation}
\delta=\delta_L+f_{NLT}(\delta_L^2-\langle\delta_L^2\rangle).
\label{eq:nonG}
\end{equation}

To obtain constraints on this parameter, we construct an 
estimator for $f_{NLT}$ 
\begin{equation}
\tilde f_{NLT} =
\frac{\zeta(\Delta)/2}
{\xi_L(\theta_1)\xi_L(\theta_2) + 
\xi_L(\theta_1)\xi_L(\theta_3) + \xi_L(\theta_3)\xi_L(\theta_2)}
\end{equation}
where $\zeta$ is our measurement in data or simulation maps.
We calculated the two-point correlation function $\xi_L$ 
analytically, to avoid any bias from the non-linear construction
\citep[c.f.][]{SzapudiEtal1999a}.
We used the same best fit power spectrum as for the simulations,
as well as taking into account beam and pixel window functions.
Since previous measurements already established the weakness of
non-Gaussianity, our Gaussian simulations should be accurate enough
to calculate the variance \citep{KomatsuEtal2003a}. 
Applying the same estimator to our 100 simulations, we obtained
error bars for $f_{NLT}$ estimated from each particular configuration.

The simplicity of the phenomenological model lies in the fact
that a constant value of $f_{NLT}$ is assumed. We do not attempt
to combine our estimates optimally, instead we use simple
considerations. 
The signal increases towards small scales in this model, while
noise dominates on the smallest scales. Since we already discarded
the smallest scales when using $N_{size}=256$, it is intuitively
clear that most signal pertaining to this model 
will be concentrated in the small fraction of the triangles 
corresponding to small scales, in particular the skewness. 
To confirm this we generated
and analyzed a set of  non-Gaussian simulations 
according to Equation~\ref{eq:nonG}, with $f_{NLT}$ equal to 1000, 2000, 
3000, 4000, and 6000. Inspection of the configurations together
with the errorbars from the Gaussian simulations confirmed the
above idea. Therefore we decided to use the skewness, which
corresponds to giving zero weight to all other configurations
when combining our $f_{NLT}$ estimators. From these we obtain
\begin{equation}
f_{NLT} \sim -110\pm 150,
\end{equation}
where the errorbar was estimated from the Gaussian simulations.
\cite{GaztanagaWagg2003} quotes similar constraints for
a low quadrupole CDM model, but their errorbars are
a factor of two larger for a $\Lambda$CDM model similar to the
one we use. 
The fact that we obtained a factor of two tighter constraints than
\cite{GaztanagaWagg2003} suggests
that using cross-correlations is superior to auto-correlations
for three-point statistics of WMAP 

\begin{figure}[htb]
\epsscale{1.}
\plotone{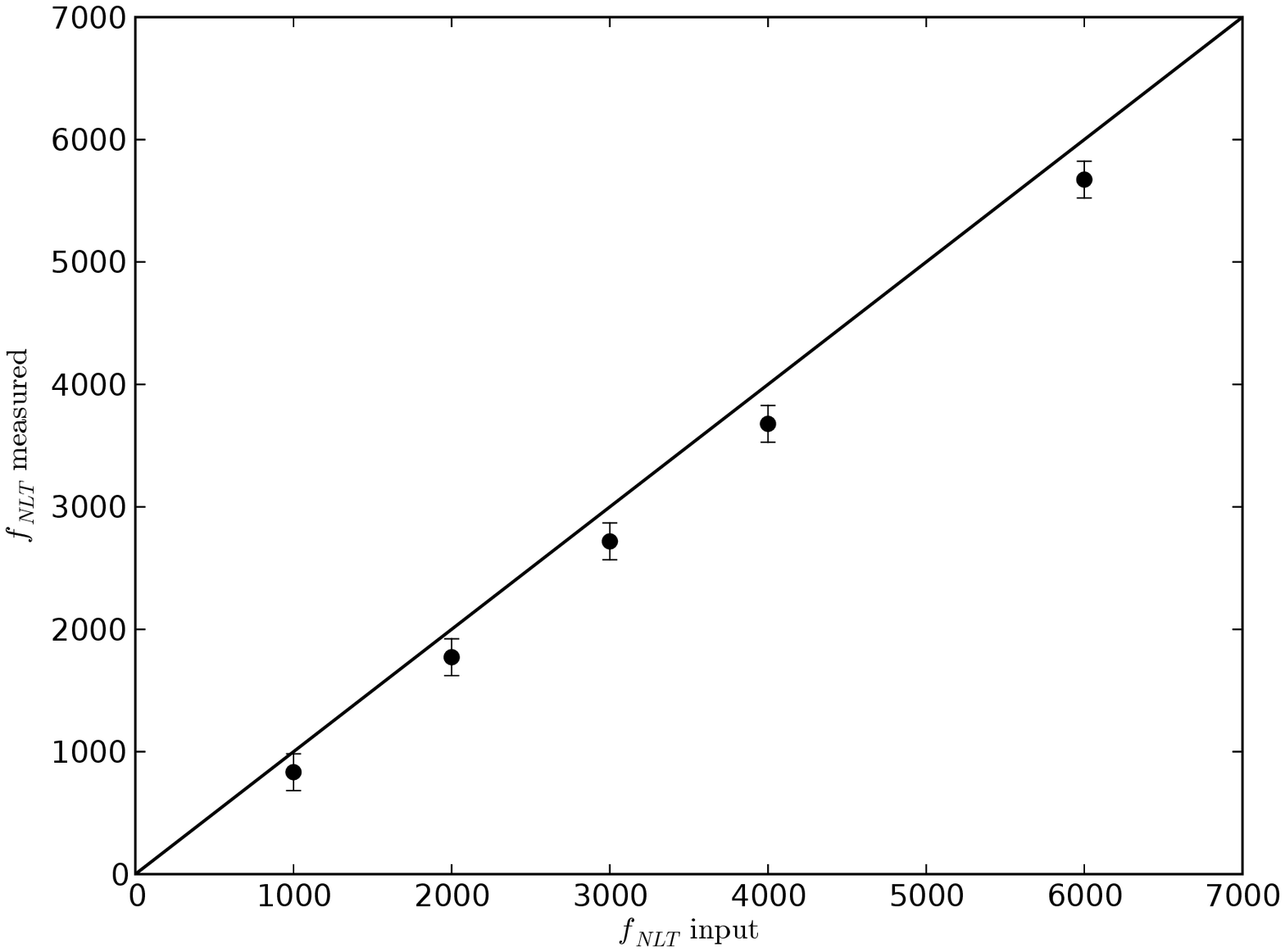}
\caption{Comparing the measured $f_{NLT}$ values, using skewness, with 
the input $f_{NLT}$ values for the non-Gaussian simulations.
The straight line represents the input values. The points with error bars
are the measured values. All the error bars are the same value from 
100 Gaussian simulations.
}
\label{fig:fNLT}
\end{figure}

As a sanity check, we calculated the mean value of the skewness estimator
for 100 Gaussian simulations; it yields about  $f_{NLT}\simeq -8.0$. 
On the other hand, we also demonstrate that we can recover $f_{NLT}$
from the non-Gaussian simulations. All the simulations
have the same underlying Gaussian signal and noise, the only difference
is the value of $f_{NLT}$. According to Figure~\ref{fig:fNLT} the
errors might be underestimated when $f_{NLT}>2000$, and/or there
might be a small low bias, but it is clear from the figure that
we could detect non-Gaussianity if it were present. 

A suboptimal
combination of $f_{NLT}$ estimates from all configurations weighted by their 
inverse variance yields about $f_{NLT} \sim -450\pm 500$, a significantly
weaker result, confirming the intuitive idea that most of the signal
is concentrated on small scales.

Some of the results in this paper have been derived using 
the HEALPix \citep{GorskiEtal1999} package.
We acknowledge the use of the Legacy Archive 
for Microwave Background Data Analysis (LAMBDA). 
Support for LAMBDA is provided by the NASA Office of Space Science.
The authors were supported by NASA through AISR NAG5-11996, 
and ATP NASA NAG5-12101 as well as by
NSF grants AST02-06243, AST-0434413  and ITR 1120201-128440.





\end{document}